\documentclass[superscriptaddress,aps,prl,floats,twocolumn,twoside,
floatfix,showpacs]{revtex4}

\usepackage{amsmath}
\usepackage{amsfonts}
\usepackage{amssymb}
\usepackage{graphicx}
\usepackage{bm}
\usepackage{hyperref}

\begin{document}

\title{
Multifractality and Conformal Invariance at 2D Metal-Insulator
Transition in the Spin-Orbit Symmetry Class}

\author{H. Obuse}
\affiliation{Condensed Matter Theory Laboratory, RIKEN, Wako,
Saitama 351-0198, Japan}
\author{A. R. Subramaniam}
\affiliation{James Franck Institute, University of Chicago, 5640
South Ellis Avenue, Chicago, Illinois 60637, USA}
\affiliation{Kavli Institute for Theoretical Physics, University of
California, Santa Barbara, California 93106, USA}
\author{A. Furusaki}
\affiliation{Condensed Matter Theory Laboratory, RIKEN, Wako,
Saitama 351-0198, Japan}
\author{I. A. Gruzberg}
\affiliation{James Franck Institute, University of Chicago, 5640
South Ellis Avenue, Chicago, Illinois 60637, USA}
\affiliation{Kavli Institute for Theoretical Physics, University of
California, Santa Barbara, California 93106, USA}
\author{A. W. W. Ludwig}
\affiliation{Kavli Institute for Theoretical Physics, University of
California, Santa Barbara, California 93106, USA}
\affiliation{Physics Department, University of California, Santa
Barbara, California 93106, USA}

\begin{abstract}
We study the multifractality (MF) of critical wave functions at
boundaries and corners at the metal-insulator transition (MIT) for
noninteracting electrons in the two-dimensional (2D) spin-orbit
(symplectic) universality class. We find that the MF exponents near
a boundary are different from those in the bulk. The exponents at a
corner are found to be directly related to those at a straight
boundary through a relation arising from conformal invariance. This
provides direct numerical evidence for conformal invariance at the
2D spin-orbit MIT. The presence of boundaries modifies the MF of the
whole sample even in the thermodynamic limit.
\end{abstract}


\date{\today}

\maketitle

Anderson metal-insulator transitions (MITs), i.e.,
localization-delocalization transitions of noninteracting electrons,
are continuous phase transitions driven by disorder. At the
transition, wave functions (WFs) are neither localized nor simply
extended, but are complicated scale invariant fractals exhibiting
multifractal behavior characterized by a continuous set of scaling
exponents \cite{Wegner,Castellani,Janssen,Mirlinreview}.
In two dimensions (2D) the universal critical properties at a host
of conventional phase transitions are known to be described by
conformal field theories (CFTs) \cite{BPZ}. It is natural to expect that
disorder-averaged observables at a localization transition in 2D are
also governed by a CFT. If so, then conformal symmetry should impose
severe constraints on averages of local quantities, including
moments of WF amplitudes.

In a recent
Letter Subramaniam \textit{et al.}\ \cite{Subramaniam2006} extended
the notion of multifractality (MF) to the boundaries of the sample
(``surface MF''), and showed that near boundaries critical WFs are
characterized by MF exponents that are different from those in the
bulk. Moreover, it was predicted that the MF of the entire system
depends crucially on the presence or absence of boundaries, even in
the thermodynamic limit.
In this Letter we study surface MF at the
MIT in 2D for non-interacting electrons with
spin-orbit scattering (symplectic universality class) \cite{Hikami},
and extend the surface MF analysis to boundaries with corners
(``corner MF'').
Conformal symmetry, if present, would lead (following
\cite{Cardy1984}) to a simple exact prediction relating
corner and surface MF exponents.
Here, we show numerically that at
the 2D spin-orbit MIT this prediction is indeed valid, thereby
providing direct evidence for the presence of conformal symmetry at
this MIT. We also confirm the dependence of the MF of the whole
system on the presence of boundaries, as predicted in
\cite{Subramaniam2006}.

\begin{figure}
\centering
\includegraphics[width=0.45\textwidth]{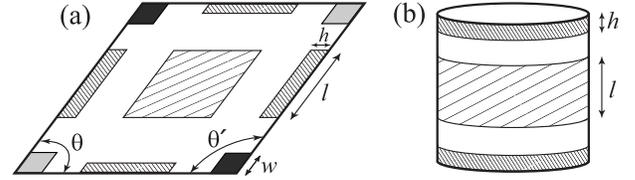}
\caption{Systems studied: (a) A rhombus with the
bulk, surface and corner regions of sizes 
$l \times l$, 
$l \times
h$, and 
$w \times w $ 
sites, correspondingly; (b) A cylinder with
the bulk ($L \times l$) and surface ($L \times h$) regions. }
\label{fig:1}
\end{figure}

We begin by introducing corner and surface MF
\cite{Subramaniam_unpublished,Subramaniam2006} for a rhombus
[Fig.~\ref{fig:1}(a)] and a cylinder [Fig.~\ref{fig:1}(b)], both
having edges of length $L$. All WFs $\psi(\boldsymbol{r})$ vanish at
the boundaries. We define the corner ($\theta$) with opening angle
$\theta$, surface (s), and bulk (b) regions of the rhombus, and
similar regions s, b of the cylinder, as illustrated in
Fig.~\ref{fig:1}. In each region $\theta$, s, or b, the  MF of WFs
is characterized by the scaling of the moments of
$|\psi(\boldsymbol{r})|^2$ with the system size $L$ (all WFs
$\psi(\boldsymbol{r})$ are normalized),
\begin{align}
L^{d_\mathrm{x}} \overline{ |\psi(\boldsymbol{r})|^{2q}} &\sim
L^{-\tau_q^\mathrm{x}}, & (\mathrm{x}=\theta, \mathrm{s},
\mathrm{b}, \mathrm{w}), \label{tau_q}
\end{align}
where $d_\mathrm{x}$ is the spatial dimension of each region
($d_\mathrm{b}=2$, $d_\mathrm{s}=1$, and $d_\theta=0$). The overbar
represents the ensemble (disorder) average and the simultaneous
spatial average over a region x surrounding the point
$\boldsymbol{r}$. $\tau_q^\mathrm{b}$, $\tau_q^\mathrm{s}$, and
$\tau_q^\theta$ are the bulk, surface, and corner MF exponents,
respectively. By $\mathrm{x} = \mathrm{w}$ we label quantities
computed by spatially averaging over the whole system
($d_\mathrm{w} = 2$) \cite{Subramaniam2006}.

Nonvanishing anomalous dimensions $\Delta_q^\mathrm{x}$,
\begin{equation}
\Delta_q^\mathrm{x} \equiv \tau_q^{\mathrm{x}} - 2q + d_\mathrm{x},
\label{Delta_q}
\end{equation}
distinguish a critical point from a simple metallic phase in which
$\Delta_q^\mathrm{x} \equiv 0$. By the definition (\ref{tau_q},
\ref{Delta_q}), $\Delta_q^\mathrm{x}$ vanish at $q=0$ and 1. The
exponent $\mu$ defined in \cite{Subramaniam2006} is absent in
Eq.\ (\ref{Delta_q}) because the local density of states is independent
of energy at the spin-orbit MIT.
The MF singularity spectra $f^\mathrm{x}(\alpha)$ are obtained from
$\tau^\mathrm{x}_q$ by Legendre transformation,
\begin{align}
f^\mathrm{x}(\alpha^\mathrm{x}) &= \alpha^\mathrm{x} q -
\tau^\mathrm{x}_q, & \alpha^\mathrm{x} = d\tau^\mathrm{x}_q/dq.
\label{Legendre}
\end{align}
$f^\mathrm{x}(\alpha)$ have the meaning of fractal dimensions: the
number of points $\boldsymbol{r}\in\mathrm{x}$, where
$|\psi(\boldsymbol{r})|^2$ scales as $L^{-\alpha}$, is proportional
to $L^{f^\mathrm{x}(\alpha)}$. This gives a direct relation between
$f^\mathrm{x}(\alpha)$ and the distribution functions of WF
amplitudes:
\begin{align}
P_\mathrm{x}(|\psi|^2) &\sim |\psi|^{-2} L^{f^\mathrm{x}(\alpha) -
d_\mathrm{x}}, & \alpha = -\ln |\psi|^2/\ln L. \label{distribution}
\end{align}
Since $f^\mathrm{\theta}(\alpha^\theta) \leqslant d_\theta= 0$, the
ensemble average is essential \cite{Evers2000} for defining corner
MF.

Suppose that the $q$-th moment
$\overline{|\psi(\boldsymbol{r})|^{2q}}$ is represented by a local
operator in an underlying critical field theory describing disorder
averages. The scaling dimension of this operator will then equal
$\Delta_q$ \cite{Duplantier}. If the field theory possesses
conformal invariance and if the operator is (Virasoro) primary, then
the relation $\Delta_q^{\theta} = \frac{\pi}{\theta}
\Delta_q^\mathrm{s}$ between the surface and corner exponents can be
derived \cite{Cardy1984} from the conformal mapping
$w=z^{\theta/\pi}$. This yields
\begin{align}
\alpha_q^{\theta}-2 &= \frac{\pi}{\theta}(\alpha_q^\mathrm{s}-2), &
f^\theta(\alpha_q^\theta) &= \frac{\pi}{\theta} \left[
f^\mathrm{s}(\alpha_q^{\text{s}}) - 1 \right]. \label{alpha^theta}
\end{align}
The validity of these relations provides direct evidence for
conformal invariance at a 2D localization critical point, and for 
the primary nature of this operator. We note,
however, that Eqs.\ (\ref{alpha^theta}) are valid only if
$\alpha^\theta_q>0$, because $\alpha$ is non-negative for normalized
WFs \cite{Janssen, Mirlinreview}. It is expected \cite{Evers2000}
that for $q>q_\theta$ (where $q_\theta$ is a solution to
$\alpha_q^\theta=0$) the exponents $\tau_q^\theta$ become
independent of $q$, while $\alpha^\theta_q=0$ [Eq.\
(\ref{Legendre})]. With the definition (\ref{Delta_q}) this leads to
a modified relation between $\Delta_q^\theta$ and
$\Delta_q^\mathrm{s}$:
\begin{equation}
\Delta_q^\theta = \left\{
\begin{aligned}
&  \dfrac{\pi}{\theta} \Delta_{q}^\mathrm{s}, & q < q_\theta, \\
&  \dfrac{\pi}{\theta} \Delta_{q_\theta}^\mathrm{s}-2(q-q_\theta),
& q > q_\theta.
\end{aligned}
\right. \label{Delta_q^theta-modified}
\end{equation}

In \cite{Subramaniam2006} it was argued that when the MF in the
whole sample with a smooth boundary is analyzed, the lowest of the
$\tau_q$ exponents for bulk and boundary dominates:
$\tau_q^\mathrm{w} = \min(\tau_q^\mathrm{b}, \tau_q^\mathrm{s})$.
Points where the curves $\tau_q^\mathrm{b}$ and $\tau_q^\mathrm{s}$
intersect translate into linear segments on the plot of
$f^\mathrm{w}(\alpha)$ (necessarily convex), interpolating between
$f^\mathrm{b}(\alpha)$ and $f^\mathrm{s}(\alpha)$, see Fig.\ 3 in
\cite{Subramaniam2006}.

In a recent paper Mirlin \textit{et al.}\ pointed out that
$\Delta_q^\mathrm{b}$ obey the relation
$\Delta_q^\mathrm{b}=\Delta_{1-q}^\mathrm{b}$, which is expected to
hold also for surface MF \cite{Mirlin2006}. In two dimensions this
leads to
\begin{align}
f^\mathrm{x}(\alpha^\mathrm{x}_{1-q})-\frac{\alpha^\mathrm{x}_{1-q}}{2}
& = f^\mathrm{x}(\alpha^\mathrm{x}_q) -
\frac{\alpha^\mathrm{x}_q}{2}, &
\alpha^\mathrm{x}_{1-q}=4-\alpha^\mathrm{x}_q,
\label{f(alpha)-relation}
\end{align}
and implies that $\alpha^\mathrm{x}$ cannot exceed 4.

To test all these theoretical predictions for the 2D spin-orbit MIT,
we employ the ``SU(2) model'' defined in \cite{Asada2002}, which is
a tight-binding model on a 2D square lattice with on-site disorder
and a random SU(2) nearest-neighbor hopping. We consider four
different lattice geometries: (i) {\it torus}, i.e., a square
lattice with periodic boundary conditions (PBC) imposed in the $x$
and $y$ directions, (ii) {\it cylinder} [Fig.\ 1(b)] with PBC
imposed in the $x$ direction and open boundary conditions (OBC) in
the  $y$ direction, (iii) {\it square} with OBC in the $x$ and $y$
directions, and (iv) {\it rhombus} [Fig.\ 1(a)] with $\theta=\pi/4$
and $\theta'=3\pi/4$ and OBC imposed in the $x$ and $y$ directions.
In all these geometries the number of lattice sites is $L^2$.

For the scaling analysis the system size $L$ is varied through
$L=24, 30, 36, \ldots, 120$. For a fixed on-site disorder strength
$W_c$, we examined $6\times10^4$ samples with different disorder
configurations for each $L$. We have used the forced oscillator
method \cite{Nakayama2001} to diagonalize the Hamiltonian, and
extracted one critical WF from each sample which had the energy
eigenvalue closest to the critical energy $E_c=1$ at $W_c=5.952$ (in
the unit of hopping strength). For the results presented below, we
have set $l=L/6$, $h=1$ for the cylinders [Fig.\ \ref{fig:1}(b)],
and $w=4$ for the corners [Fig.\ \ref{fig:1}(a)]. We have
numerically confirmed that the exponents computed in the bulk
regions of rhombi and cylinders agree with those of tori within
statistical error bars. Also, the MF exponents for the surface
region of rhombi are, within error bars, equal to those computed for
the surface region of cylinders. In the following figures the bulk
(surface) exponents are those computed for tori (for the surface
region of cylinders).

\begin{figure}
\centering
\includegraphics[width=0.45\textwidth]{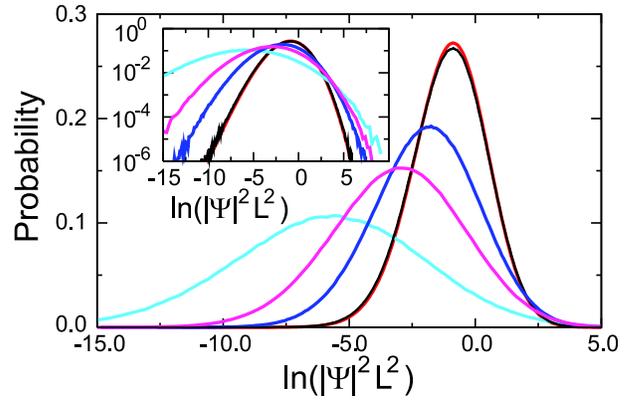}
\caption{
PDFs
of
logarithm of WF amplitudes
on tori (red), in the
bulk region of cylinders (black), in the surface region of cylinders
(blue), and in the corner region with $\theta=\pi/2$ (pink) and
$\theta=\pi/4$ (light blue); $L=120$. Inset: Semi-logarithmic plot. }
\label{fig:2}
\end{figure}

Figure \ref{fig:2} shows the probability distribution functions 
(PDFs)
of
$\ln|\psi(\boldsymbol{r})|^2$ measured for $\boldsymbol{r}$ at
corners with angle $\theta=\pi/4$ (light blue) and $\theta=\pi/2$ (pink),
at the boundary of cylinders (blue), and in the bulk region of
cylinders (black) at the fixed
$L=120$. Each
PDF
is normalized in the region where it is
defined. The 
PDF
calculated for tori is also shown
in red, and it agrees quantitatively with the bulk 
PDF
(black), as expected. Clearly, the 
PDFs
for bulk, surface, and corner with $\theta=\pi/2$ and
$\theta=\pi/4$, 
are all different, and, in this order, the peak
position is shifted to the left, in agreement with the expectation
that WF amplitudes should be smaller near edges. In the same order,
the distributions become broader with longer (presumably power-law)
tails at $|\psi|^2L^2\gg1$. This means that for large $q$ the
moments $\overline{|\psi(\boldsymbol{r})|^{2q}}$ can become larger
near edges (corners) than in the bulk, as the higher moments are
dominated by long tails \cite{Evers2000}.

\begin{figure}
\centering
\includegraphics[width=0.45\textwidth]{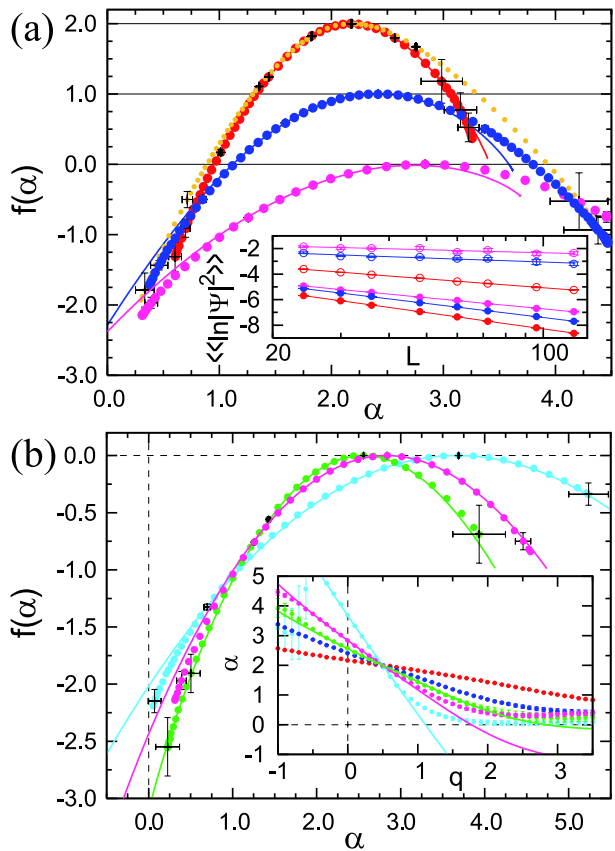}
\caption{ (a) Bulk (red), surface (blue), corner
$\theta=\pi/2$ (pink), and whole cylinder (orange) $f(\alpha)$
spectra, with error bars shown at integer values of $q$. Red, blue,
and pink curves represent $f^\mathrm{x}(4-\alpha)+\alpha-2$. Inset:
Scaling plot of Eq.\ (\ref{alpha}) at $q=1$ (filled circles) and
$q=3$ (open circles). (b) Corner $f(\alpha)$ spectra at
$\theta=\pi/4$ (light blue), $\pi/2$ (pink), and $3\pi/4$ (green), with
error bars shown at integer $q$ (and at $q=-0.5$ for
$\theta=\pi/4$). Curves represent the conformal relation
(\ref{alpha^theta}). Inset: Numerical results for
$\alpha^\mathrm{x}_q$ compared with Eqs.\ (\ref{alpha^theta})
(colored curves). } \label{fig:f(alpha)}
\end{figure}

We numerically obtain $\alpha_q^\mathrm{x}$ and
$f^\mathrm{x}(\alpha^\mathrm{x}_q)$ from [see (\ref{tau_q},
\ref{Legendre})]
\begin{align}
\langle\langle \ln |\psi|^2 \rangle\rangle_q & \equiv
\frac{\overline{|\psi(\boldsymbol{r})|^{2q}
\ln|\psi(\boldsymbol{r})|^2
}} {\overline{
|\psi(\boldsymbol{r})|^{2q}}} \sim - \alpha_q^\mathrm{x}\ln L,
\label{alpha}\\
\ln\overline{|\psi(\boldsymbol{r})|^{2q}} & \sim\left[
f^\mathrm{x}(\alpha_q^\mathrm{x}) - \alpha_q^\mathrm{x}q -
d_\mathrm{x} \right]\ln L . \label{f(alpha)}
\end{align}
The inset of Fig.\ \ref{fig:f(alpha)}(a) shows
$\langle\langle\ln|\psi|^2\rangle\rangle_q$ as functions of $L$,
computed for tori ($\mathrm{x} = \mathrm{b}$), at the boundary of
cylinders ($\mathrm{x} = \mathrm{s}$), and at the corners
($\theta=\pi/2$) of squares. This inset exhibits distinct scaling
behavior for bulk, surface, and corner regions for the displayed
values of $q=1,3$.

Figure \ref{fig:f(alpha)}(a) shows $f^\mathrm{x}(\alpha)$ of the
bulk, surface, and corner ($\theta=\pi/2$) regions. Clearly, in this
order, the spectra $f^\mathrm{x}(\alpha)$ are seen to become broader
and their maxima $\alpha_0^\mathrm{x}$ are shifted to the right
($\alpha_0^\mathrm{b}=2.173\pm0.001$,
 $\alpha_0^\mathrm{s}=2.417\pm0.002$,
 $\alpha_0^{\pi/2}=2.837\pm0.003$),
in accordance with Fig.\ 2 and Eq.\ (\ref{distribution}). (Recall
that the maximal values of $f^\mathrm{x}(\alpha)$ are the spatial
dimensions $d_\mathrm{x}$.) The plot of $f^\mathrm{w}(\alpha)$ for
the whole cylinder (Fig.\ \ref{fig:f(alpha)}(a), orange) clearly
represents the convex hull of $f^\mathrm{b}(\alpha)$ and
$f^\mathrm{s}(\alpha)$ \cite{Subramaniam2006, interpolation}. Notice
that $f^\mathrm{w}(\alpha)$ deviates from $f^\mathrm{b}(\alpha)$
already at $f(\alpha) \approx 1.5$ (for $q < 0$). This confirms the
prediction of \cite{Subramaniam2006} that the presence of boundaries
drastically affects the MF of the system even in the thermodynamic
limit and also in a typical sample (where $f(\alpha) \geqslant 0$
\cite{Evers2000}).

The data points of the bulk spectrum $f^\mathrm{b}(\alpha)$ [red
dots in Fig.\ \ref{fig:f(alpha)}(a)] lie on top of the  red curve
representing data points \cite{DataContinuousCurve} for
$f^\mathrm{b}(\alpha^\mathrm{b}_{1-q})$
in Eq.\ (\ref{f(alpha)-relation}).
This confirms Eq.\ (\ref{f(alpha)-relation}) for the bulk
which is the consequence of the symmetry relation \cite{Mirlin2006}.
Incidentally, the value of the typical bulk  exponent
$\alpha_0^\mathrm{b}$ agrees with earlier calculations
\cite{Schweitzer,Merkt} but not with \cite{Markos}. Furthermore, our
$\alpha_0^\mathrm{b}$ satisfies
$\pi(\alpha_0^\mathrm{b}-2)=1/\Lambda_c$ \cite{Janssen}, where
$\Lambda_c=1.843$ is the quasi-1D localization length at the MIT,
normalized by the wire width, as obtained in \cite{Asada2002}. The
surface spectrum $f^\mathrm{s}(\alpha)$ (blue) is also seen to
satisfy the relation (\ref{f(alpha)-relation}) for
$1\lesssim\alpha^\mathrm{s}\lesssim 3$, but there are discrepancies
between the blue dots and the curve
$f^\mathrm{s}(\alpha^\mathrm{s}_{1-q})$ when $\alpha^\mathrm{s}_q>3$
($q<-0.7$) and $\alpha^\mathrm{s}_q<1$ ($q>2$).
Moreover it appears that
$\alpha^\mathrm{s}_q$ can exceed 4, in contrast to
$\alpha^\mathrm{b}_q < 4$. This may question the validity of the
symmetry relation of  \cite{Mirlin2006} for boundaries, but we feel
that computations on even larger system sizes and numbers of samples
are necessary for drawing a definitive conclusion.

Figure \ref{fig:f(alpha)}(b) shows the corner spectra
$f^\theta(\alpha)$ at $\theta=3\pi/4$ (green), $\pi/2$ (pink),  and
$\pi/4$ (light blue). As $\theta$ decreases, the peak position moves to
the right ($\alpha_0^{3\pi/4}= 2.558\pm0.003$,
$\alpha_0^{\pi/2}=2.837\pm0.003$,
and $\alpha_0^{\pi/4}=
3.689\pm0.006$) and the spectra become broader, indicating that at
smaller $\theta$ the typical value of a WF amplitude is smaller but
its distribution is broader. The numerical data (dots) are compared
with the curves predicted from conformal invariance, Eq.\
(\ref{alpha^theta}), using $f^\mathrm{s}(\alpha^\mathrm{s})$ of
Fig.\ \ref{fig:f(alpha)}(a) within the range
$1\lesssim\alpha^\mathrm{s}\lesssim3$, where $|q|$ is sufficiently
small to ensure good numerical accuracy. The agreement between the
numerical data and the predicted curves is excellent, confirming the
presence of conformal symmetry.

The inset of Fig.\ \ref{fig:f(alpha)}(b) shows $\alpha^\mathrm{x}_q$
where the  curves represent $\alpha^\theta_q$
computed with $\alpha^\mathrm{s}_q$ as input in Eq.\
(\ref{alpha^theta}). Note that $\alpha^\mathrm{x}=2$ at $q=1/2$ as a
consequence of Eq.\ (\ref{f(alpha)-relation}). We see that the
numerical data for $\alpha^\theta_q$ deviate from the predicted
curves, Eq.\ (\ref{alpha^theta}), when $\alpha^\theta_q \lesssim 1$,
in order to satisfy
the constraint 
$\alpha^\theta_q>0$. We expect
that in the limit $L\to\infty$, $\alpha^\theta_q$ be given by Eq.\
(\ref{alpha^theta}) for $q<q_\theta$ and by $\alpha^\theta_q=0$ for
$q>q_\theta$.

We note that the numerical results for $\alpha^\theta_q$ exceed 4
when $q \lesssim -0.1$ for $\theta = \pi/4$, and $q \lesssim -0.7$
for $\theta = \pi/2$ [Fig.\ \ref{fig:f(alpha)}(b) and inset]. Even
the maximum $\alpha_0^\theta$ of $f^\theta(\alpha)$ will exceed 4
for sufficiently small angles $\theta$. On one hand, the maximum
corresponds to $q=0$ where the numerics are most accurate. On the
other hand, the maximal value $f^\theta(\alpha) = d_\theta = 0$ has
a direct physical meaning as the dimension of the WF support, and
must therefore appear on the $f^\theta(\alpha)$ curve. Thus, our
data strongly indicate that the symmetry relation of
\cite{Mirlin2006} is violated for corners.

\begin{figure}[t]
\centering
\includegraphics[width=0.45\textwidth]{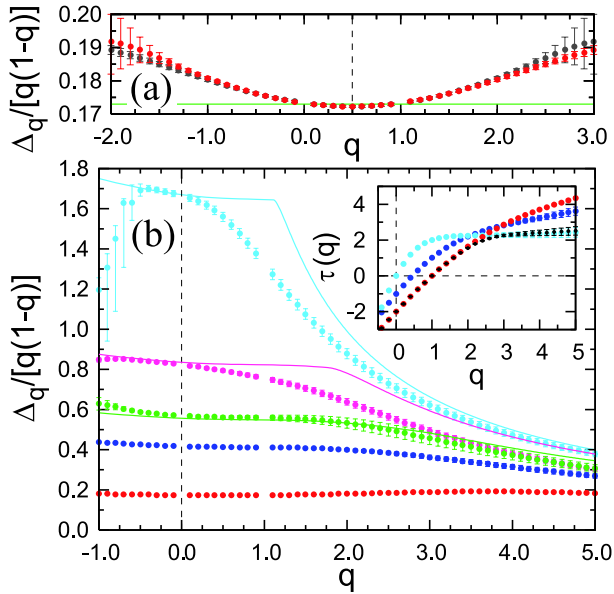}
\caption{ (a) The numerical data for the bulk
exponents $\Delta^\mathrm{b}_q$ (red) are compared with their mirror
image $\Delta^\mathrm{b}_{1-q}$ (grey). The green line represents
$\alpha_0^\mathrm{b}-2$. (b) The exponents $\Delta_q/[q(1-q)]$ for
the bulk (red), the surface (blue), and corners with $\theta=\pi/4$
(light blue), $\pi/2$ (pink), and $3\pi/4$ (green). The curves represent
the theoretical prediction, Eq.\ (\ref{Delta_q^theta-modified}).
Inset: Bulk (red), surface (blue), and corner ($\theta=\pi/4$, light blue)
exponents $\tau^\mathrm{x}_q$, and $\tau^\mathrm{w}_q$ for a whole
rhombus with $\theta=\pi/4$ (black).} \label{fig:Delta}
\end{figure}

The anomalous dimensions $\Delta_q^\mathrm{x}$ are computed
numerically from $\overline{|\psi(\boldsymbol{r})|^{2q}}/
{(\;\!\overline{|\psi(\boldsymbol{r})|^2}\;\!)}^q \sim
L^{-\Delta_q^\mathrm{x}}$, which follows from Eqs.\ (\ref{tau_q})
and (\ref{Delta_q}). Figure \ref{fig:Delta}(a) shows the bulk
anomalous dimension $\Delta_q^\mathrm{b}$ (red) and its mirror image
across the $q = 1/2$ line $\Delta_{1-q}^\mathrm{b}$ (grey), both
rescaled by $q(1-q)$. Note that this rescaling magnifies small
numerical errors around $q=0$ and $q=1$. Nevertheless the numerical
data satisfy the relation
$\Delta^\mathrm{b}_q=\Delta_{1-q}^\mathrm{b}$ of \cite{Mirlin2006}
for $-1<q<2$ where statistical errors are small. It is also clear
from Fig.\ \ref{fig:Delta}(a) that $\Delta_q^\mathrm{b}/[q(1-q)]$
varies with $q$, which means that the bulk spectrum
$f^\mathrm{b}(\alpha)$ is not exactly parabolic.

Figure \ref{fig:Delta}(b) compares $\Delta_q^\mathrm{x}$ for bulk,
surface, and corners with $\theta=\pi/4, \pi/2$, and $3\pi/4$. The
solid curves represent the theoretical prediction
(\ref{Delta_q^theta-modified}) from the conformal mapping, where
$\Delta_q^\mathrm{s}$ is taken from Fig.\ \ref{fig:Delta}(b).
For sufficiently small values of $|q|$ the numerical results of
$\Delta_q^\theta$ are in good quantitative agreement with the
prediction (\ref{Delta_q^theta-modified}). It is precisely for small
$|q|$ that the numerical data are most accurate \cite{discrepancy}.
This provides direct evidence for the presence of conformal symmetry
at the 
2D spin-orbit MIT.

The inset of Fig.\ \ref{fig:Delta}(b) shows the exponents
$\tau_q^\mathrm{x}$ for bulk, surface, and corners. We see that
$\tau_q^{\pi/4}$ (light blue) is constant for $q > q_{\pi/4} \approx 1$
reflecting the exchange between top and bottom lines in Eq.\
(\ref{Delta_q^theta-modified}) which happens at
$\alpha^{\pi/4}_q=0$. It appears that $\tau^{\pi/4}_q$ becomes
smaller than both $\tau^\mathrm{s}_q$ and $\tau^\mathrm{b}_q$ for
$q\gtrsim2.5$, which is when the corner exponent $\tau^{\pi/4}_q$
controls the MF of the whole sample with a $\pi/4$ corner, as shown
in black. In a sample without corners such as a cylinder, the
surface exponent $\tau_q^s$ controls the MF of the entire sample for
sufficiently large $q$. This confirms and generalizes the
predictions made in  \cite{Subramaniam2006}.

In summary, we 
studied
bulk, surface, and corner 
multifractality
at the MIT in the 2D spin-orbit symmetry class. We
provided direct numerical evidence for the presence of conformal
symmetry at this critical point, and confirmed predictions of
\cite{Subramaniam2006} that boundaries affect MF of the whole system
even in the thermodynamic limit. We also tested the validity of the
symmetry relation of  \cite{Mirlin2006} for the bulk, surface and
corners. It appears that the relation holds in the bulk, but is
violated at corners.

We thank A. Mirlin for discussions. This work was supported by
NAREGI Grant from MEXT of Japan,
the NSF under Grant No. PHY99-07949, the NSF MRSEC under
DMR-0213745, the NSF Career award DMR-0448820, the Alfred P. Sloan
Foundation, and Research Corporation. Numerical calculations
were 
performed on the RIKEN Super Combined Cluster System.
--- {\it Note added:} Recent numerical work on
the bulk MF in the 2D symplectic class \cite{Mildenberger} agrees
with our results for the bulk.


\begin{thebibliography}{99}
\bibitem{interpolation}
The interpolating portions of $f^\mathrm{w}(\alpha)$ are slightly curved
due to finite size effects.
\bibitem{discrepancy}
For $q \gtrsim 1$, Fig.\ \ref{fig:Delta}(b) shows discrepancies
between the numerical results and the prediction from CFT, enhanced
by the $(1-q)^{-1}$ factor. In this range of $q$ the moments of
$|\psi|^2$ are dominated by rare events, which makes it difficult to
obtain accurate numerical data.
\end{thebibliography}
\end{document}